\documentclass{article}  
\usepackage{bigsky2009}
\usepackage{graphicx}
\frompage{000} \topage{000}                                              
\def\rightmark{Jet Reconstruction in Heavy Ion Collisions}
\def\leftmark{S. Salur}
 
\title{Jet Reconstruction in Heavy Ion Collisions} 
\authors{
{Sevil Salur$^1$ %
}\\[2.812mm]
{\normalsize
\hspace*{-8pt}$^1$ Lawrence Berkeley National Laboratory, 1 Cyclotron Road MS-70R0319, Berkeley, CA 94720\\[0.2ex] 
}}
 
\abstract{Measurements of  strong suppression of inclusive hadron distributions and di-hadron correlations at high $p_{T}$, while providing evidence for partonic energy loss,  also suffer from geometric biases due to the competition of energy loss and fragmentation.  The measurements of fully reconstructed jets is expected to lack these biases as the energy flow is measured independently of the fragmentation details. In this article, we review the recent results from the heavy ion collisions collected by the STAR experiment at RHIC on direct jet reconstruction utilizing the modern sequential recombination and cone jet reconstruction algorithms together with their background subtraction techniques. In order to assess the jet reconstruction biases a comparison with the jet cross section measurement in $\sqrt{s}=200$ GeV p+p collisions scaled by the number of binary nucleon-nucleon collisions to account for nuclear geometric effects is performed.  Comparison of the inclusive jet cross section obtained in central Au+Au events with that in  $p+p$ collisions, published previously by STAR, suggests that unbiased jet reconstruction in the complex heavy ion environment indeed may be possible. }

 
\begin{document}
 
\maketitle
\setcounter{page}{1}

\section{Introduction}\label{intro}
 
Jets are remnants of hard-scattered quarks and gluons and they are studied extensively in high energy collisions of all kinds.    Despite the naive expectations, jets are not fundamental objects in Quantum Chromo Dynamics (QCD), but instead they are artificial event properties defined by hand, i.e.,  well-defined and easy to measure from the hadronic final-state, easy to calculate in pQCD from the partonic final-state and closely related to the final-state quarks and gluons  \cite{seymor}.  A well defined jet definition  allows us to study the fundamental objects of pQCD which are the quarks and the gluons.  But recently, many new areas outside the standard QCD in high energy physics  are utilizing jet physics to answer major questions.  For example at RHIC, jets are put to a new use to study the hot QCD matter through their interaction and energy loss in the medium (``jet quenching").

Direct jet measurements in $p+p$ collisions at RHIC have been carried out since the third year of RHIC operations \cite{starpp}.  However until now, to avoid the complex backgrounds of heavy ion events, inclusive hadron distributions and di-hadron correlations at high transverse momentum are utilized and indirect measurements of jet quenching have been made at RHIC.  However, these measurements of jet fragmentation particles are biased towards the population of jets that has the least interaction with the medium.  Since 2006, the STAR barrel electromagnetic calorimeter (BEMC) has been operated with full azimuthal coverage ($\phi$) and large pseudorapidity ($\eta$) acceptance. This detector upgrade together with the increased beam luminosities of RHIC and data recording capabilities of STAR, enables the study of full jet reconstruction in heavy ion collisions for the first time at RHIC \cite{salurww}. This article discusses results from a recent new approach of full jet reconstruction measurement in heavy ion collisions, utilizing the high luminosity Au+Au data set collected by the STAR experiment from 2007 RHIC run.  The experimental details can be found in \cite{me} for the direct measurement of jets and \cite{jor,bruna,elena} for the accompanying jet fragmentation studies in heavy ion collisions utilizing the STAR experiment.



\section{Jet Reconstruction Analysis}

During the last 20 years, various jet reconstruction algorithms have been developed for both leptonic and hadronic colliders.   For a detailed overview of jet algorithms in high energy collisions, see  \cite{me,davidE} and the references therein.   Here we will briefly discuss the algorithms used for the STAR analysis.  Two kinds of jet reconstruction algorithms are utillized; seeded cone (leading order high seed cone (LOHSC)) and sequential recombination ($\rm k_{T}$ and Cambridge/Aachen).

The cone algorithm is based on the simple picture that a jet consists of a large amount of hadronic energy in a small angular region. Therefore, the main method for the cone algorithm is to combine particles in $\eta - \phi $ space with their neighbors within a cone of radius R ($R=\sqrt{ \Delta \phi ^{2}+ \Delta \eta^{2} }$).   The sequential recombination algorithms  combine objects in relative to the closeness of their $p_{T}$. Particles are merged into a new cluster via successive pair-wise recombination.   In the sequential recombination algorithm, arbitrarily shaped jets are allowed to follow the energy flow resulting in less bias on the reconstructed jet shape than with the cone algorithm \cite{catchment}.  Algorithmic details of cone and sequential recombination can be found in \cite{jets,kt,ktref,blazey} and the references therein.

Most recently a new approach to jet reconstruction and background subtraction, motivated by the need of precision jet measurements in the search for new physics in high luminosity $p+p$ collisions at the LHC is developed by M. Cacciari, G. Salam and G. Soyez \cite{catchment,salamtalk}. A key feature of their approach is a new QCD inspired algorithm for separating jets from the large backgrounds due to pile up. As it turns out from simulations,  these improved techniques can also be used in heavy ion environments where the background subtraction is essential for jet measurements.  Sequential recombination algorithms ($\rm k_{T}$, anti-$\rm k_{T}$ and Cambridge/Aachen (CAMB)) encoded in the $FastJet$ suite of programs  \cite{catchment,antikt}, along with an alternative seeded cone algorithm (labeled LOHSC) are utilized to search for jets in the Au+Au collisions. A seedless infrared-safe cone algorithm (SISCone) \cite{sis} which is also available in the $FastJet$ suite of programs as a plug-in,  is already used in $p+p$ collisions at $\sqrt{s}=200$ GeV and the first results  can be found in \cite{elena}. 

\section{Results}

\begin{figure}[]
\begin{center}
\resizebox{0.60\textwidth}{!}{%
	\includegraphics{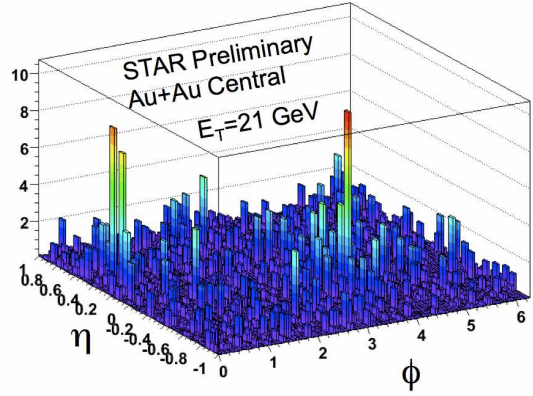} 
}	
\end{center}
\caption{21 GeV di-jet reconstructed from a central Au+Au event at $\sqrt{s_{NN}}=200$ GeV in the STAR detector \cite{me,jor}.}
\label{fig:dijets}       
\end{figure}

\begin{figure}[t!]

\centering
$\begin{array}{cc}
\resizebox{0.47\textwidth}{!}{
\includegraphics{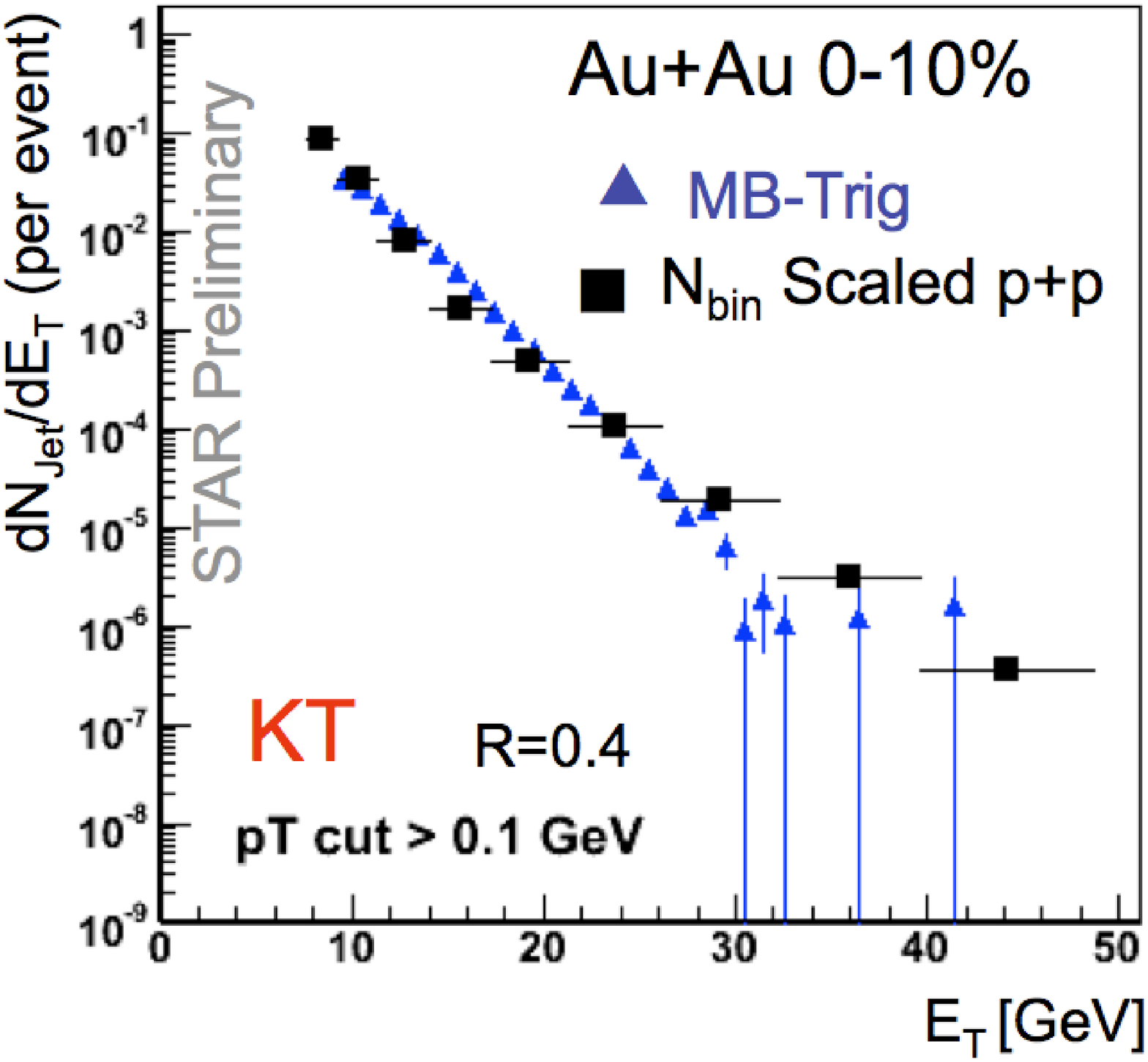}}&
\resizebox{0.47\textwidth}{!}{
\includegraphics{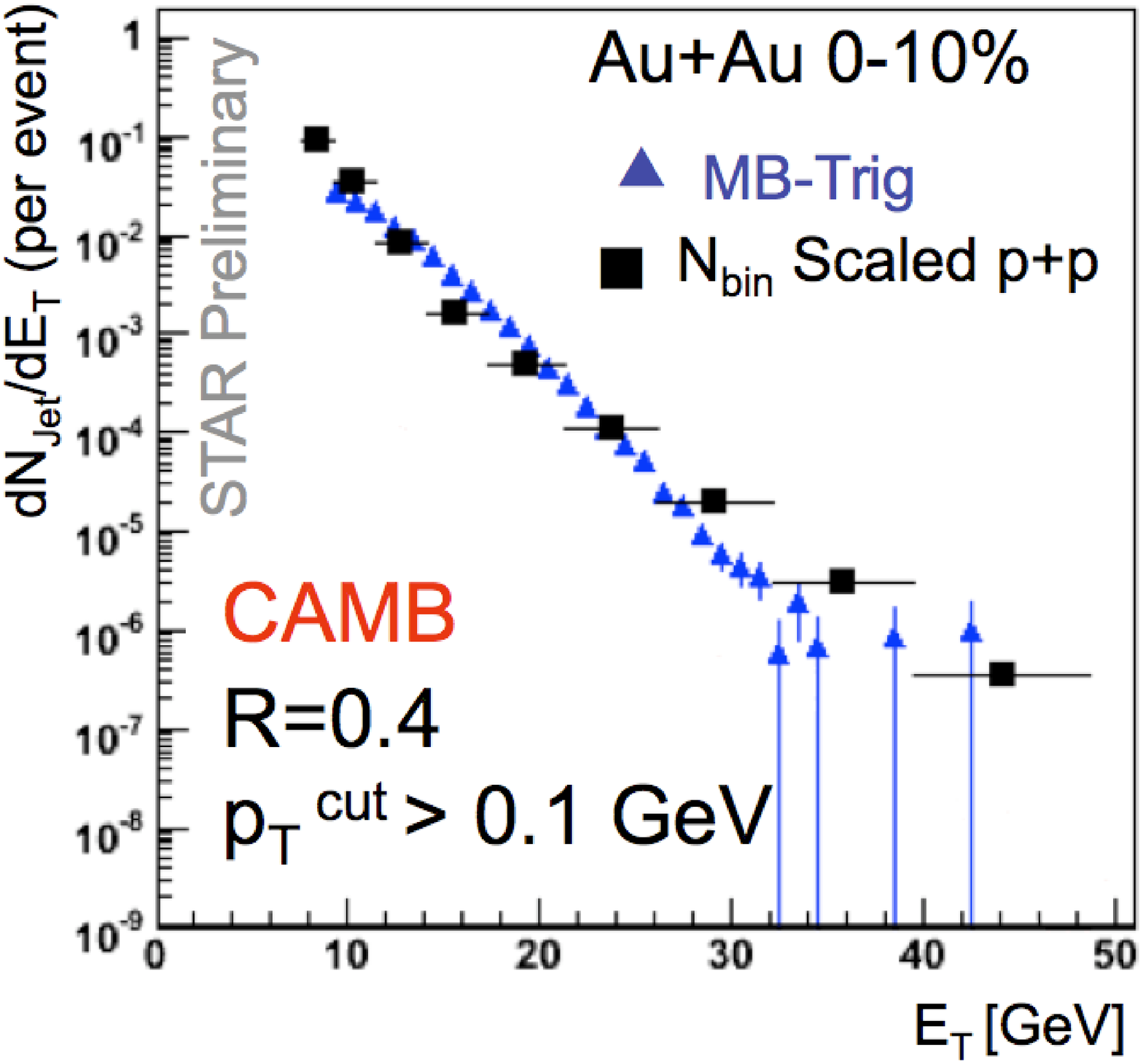}}
\\
\end{array}$

 \caption[]{
 Jet yield per event vs transverse jet energy ($E_{T}$) for the central Au+Au collisions obtained by the sequential recombination ($\rm k_{T}$, CAMB) algorithms \cite{starpp,me}. 
 Triangle symbols are from MB-Trig and corrected for efficiency, acceptance and energy resolution.  Only statistical error bars are shown for the $Au+Au$ data. Solid black squares are the distribution from $p+p$ collisions, scaled by $N_{Binary}$. The systematic uncertainty of the $p+p$ jet spectra normalization is  $\sim 50 \%$.
 
   } \label{fig:kt}
\end{figure}

\begin{figure}[t!]

\centering
\resizebox{0.49\textwidth}{!}{
\includegraphics{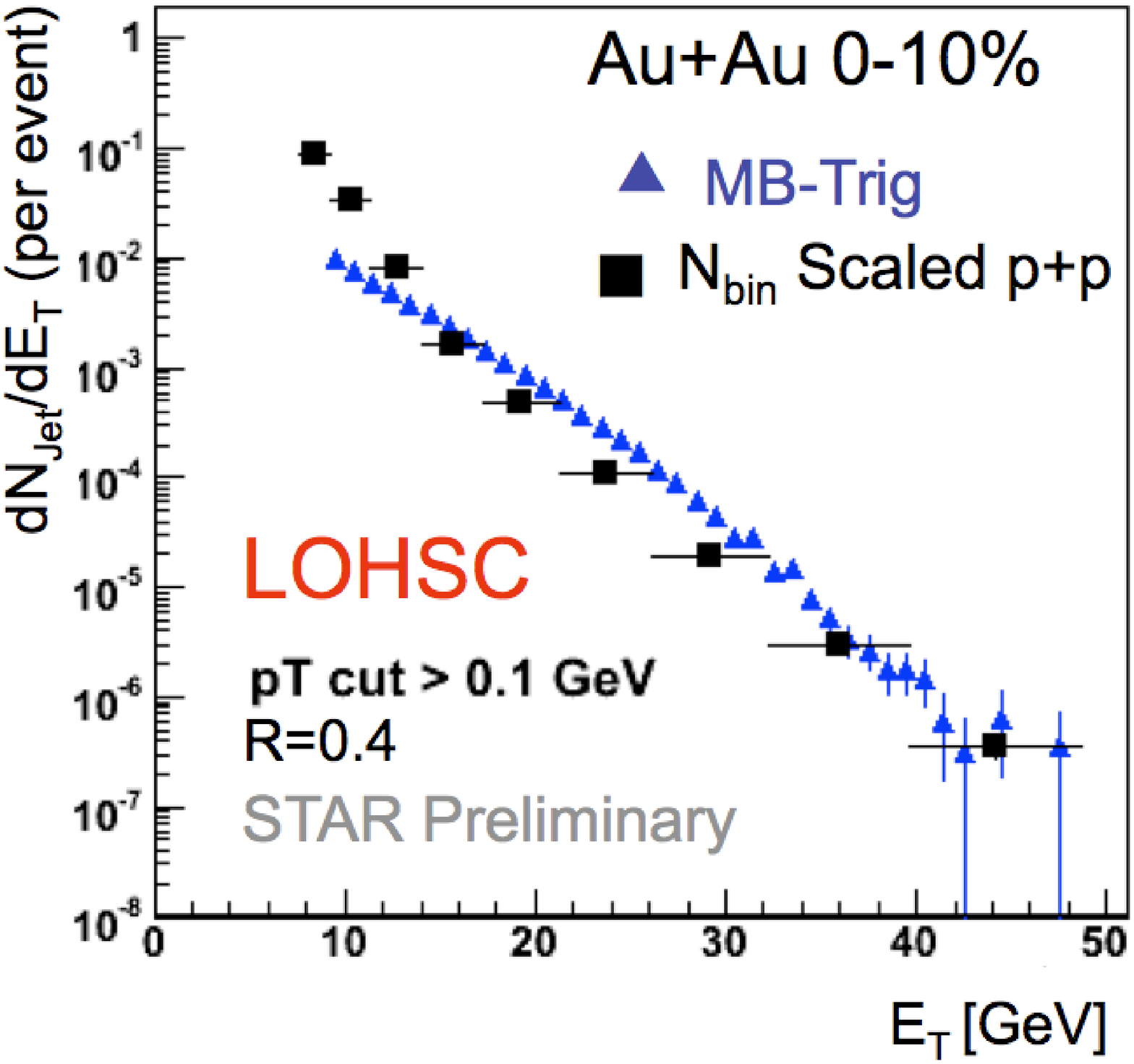}}
\\

 \caption[]{
 Jet yield per event vs transverse jet energy ($E_{T}$) for the central Au+Au collisions obtained by the Leading Order High Seed Cone (LOHSC) algorithm \cite{starpp,me}.  Triangle symbols are from MB-Trig and corrected for efficiency, acceptance and energy resolution.  Only statistical error bars are shown for the $Au+Au$ data. Solid black squares are the distribution from $p+p$ collisions, scaled by $N_{Binary}$. The systematic uncertainty of the $p+p$ jet spectra normalization is  $\sim 50 \%$.
  } \label{fig:LOHSC}
\end{figure}

Figure~\ref{fig:dijets} shows an example of an identified di-jet event for central Au+Au collisions, using both the neutral energy from the BEMC and charged particles from the Time Projection Chamber  of the STAR experiment.   In order to assess the bias of the heavy ion jet measurements, the inclusive jet cross section is compared to that from $p+p$ collisions presented in reference~\cite{starpp}.  Figure~\ref{fig:kt} shows the comparison of the inclusive jet spectrum for central Au+Au collisions (taken with a Minimum Bias online trigger ``MB-Trig'' ) to the $\rm N_{Bin}$ scaled $p+p$ spectrum, for the $\rm k_{T}$ and CAMB jet reconstruction algorithms.  Jets in $p+p$ collisions are measured the same way as in $Au+Au$ collisions, utilizing the STAR Time Projection Chamber and BEMC and correcting for missing and double counted energy \cite{starpp}. However for the $p+p$ case, a mid-point cone jet algorithm with splitting and merging steps is used.   The inclusive jet spectrum from $p+p$ collisions  agrees well with the Next-to-leading order perturbative QCD calculation \cite{nlo}.  The same comparison for the jets reconstructed with the LOHSC algorithm is presented in Figure~\ref{fig:LOHSC}. For both figures, to account for nuclear geometric effects, the $p+p$ spectrum is scaled by $\rm N_{Bin}$, the number of binary nucleon+nucleon collisions ($\rm N_{Bin}$) equivalent to a central Au+Au collisions, as calculated by a Glauber model  \cite{glauber}.


\begin{figure}[]
\begin{center}
\resizebox{0.95\textwidth}{!}{%
	\includegraphics{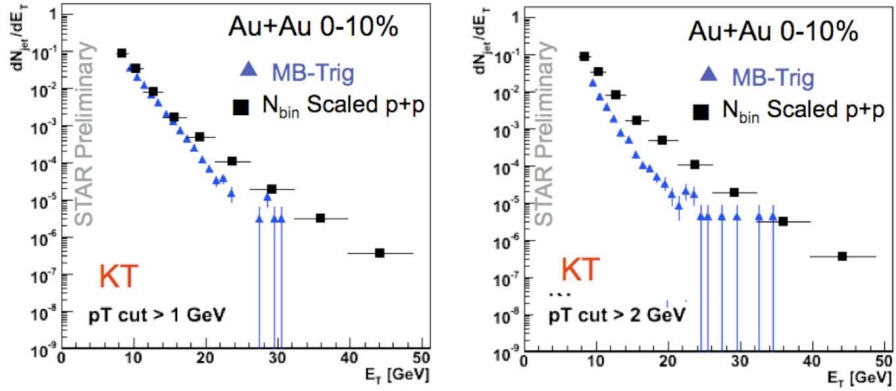} 
}	
\end{center}
\caption{Jet yield per event vs $E_{T}$ for 0-10\% central $Au+Au$ collisions obtained by the $k_{T}$  algorithm for the two selection of the $p_{T}^{cut}$. The distribution from $p+p$ collisions are scaled by  $\rm N_{Binary}$ \cite{starpp,me}.  Triangle symbols are from MB-Trig and corrected for efficiency, acceptance and energy resolution.  Only statistical error bars are shown for the $Au+Au$ data. Solid black squares are the distribution from $p+p$ collisions, scaled by $N_{Binary}$. The systematic uncertainty of the $p+p$ jet spectra normalization is  $\sim 50 \%$.}
\label{fig:panelkt}       
\end{figure}

 In the case of jet reconstruction, $\rm N_{Bin}$ scaling is expected if the reconstruction is unbiased, i.e. the jet energy is recovered independent of the fragmentation, even in the presence of strong jet quenching. This scaling is analogous to the cross section scaling of high $p_{T}$ direct photon production in heavy ion collisions, observed by the PHENIX experiment \cite{phenix}. At present, the total systematic systematic uncertainty on the normalisation of the inclusive $p+p$ jet spectrum is around 50\%. Figure~\ref{fig:kt} and Figure~\ref{fig:LOHSC} show that the heavy ion jet spectrum agrees well with the scaled $p+p$ measurement within the systematic uncertainty.  Figure~\ref{fig:panelkt} is for the jet spectra obtained with the $\rm k_{T}$ algorithm by using different threshold cuts on the track momenta and calorimeter tower energies ($p_{T}^{cut}$).  It is found that the agreement between the binary scaled $p+p$ spectra and the Au+Au measurement is worse for larger $p_{T}^{cut}$. This suggests that the threshold cuts introduce biases which are not fully corrected with the current procedure that uses fragmentation models that are developed for $e^{+}+e^{-}$ and $p+p$ collisions. It could also be an indication of modified fragmentation due to jet quenching.

\section{Conclusions}
Unbiased reconstruction of jets in central heavy ion collisions at RHIC energies would be a breakthrough to investigate the properties of the matter produced at RHIC.  The study shown here indicates that unbiased reconstruction of jets may be possible in heavy ion events. However, spectrum corrections are currently based on model calculations using PYTHIA fragmentation \cite{PYTHIA}. This aspect, together with the spectrum variations due to cuts and reconstruction algorithms, must be investigated further in order to assess the systematic uncertainties of this measurement.

Further, we utilize the reconstructed jets and study the jet shapes to test the underlying QCD theory \cite{vitev,vitev2}.  The results from the intra-jet energy distributions, jet-jet and hadron-jet correlation studies will be available in the coming months and will enable us to study the medium properties produced at RHIC. Also a copious production of very energetic jets, well above the heavy ion background is predicted to occur at the LHC \cite{peter,solan}. The large kinematic reach of high luminosity running at RHIC and at the LHC may provide sufficient lever-arm to map out the QCD evolution  of jet quenching.  The comparison of  full jet measurements in the different physical systems generated at RHIC and the LHC will provide unique and crucial insights into our understanding of jet quenching and the nature of hot QCD matter.

\section*{Acknowledgments}
We thank the organizers of 25th WWND for the opportunity and E. Bruna, W. Holzmann, G. Soyez, and I. Vitev for particularly fruitful discussions about jet reconstruction in heavy ion collisions during the workshop.

\section*{Note(s)} 
\begin{notes}
\item[a]
E-mail: ssalur@lbl.gov
\end{notes}
 
\bibliographystyle{bigsky2009}
 

\vfill\eject
\end{document}